\documentclass[a4paper, amsfonts, amssymb, amsmath, reprint, showkeys, showpacs, nofootinbib, twoside, superscriptaddress,aps,prd]{revtex4-1}
\usepackage[english]{babel}
\usepackage[utf8]{inputenc}
\usepackage{comment}
\usepackage[colorinlistoftodos, color=green!40, prependcaption]{todonotes}
\usepackage{appendix}
\usepackage{amsthm}
\usepackage{mathtools}
\usepackage{siunitx}
\usepackage{xcolor}
\usepackage{graphicx}
\usepackage{adjustbox}
\usepackage{placeins}
\usepackage[T1]{fontenc}
\usepackage{lipsum}
\usepackage{csquotes}
\usepackage{enumerate}
\usepackage{hyperref}
\usepackage{siunitx}

\usepackage{ulem}

\begin{document}

\title{Interferometric signature of higher-order images in a parametrized framework}

\author{Fabiano Feleppa}
    \email[]{ffeleppa@unisa.it}
    \affiliation{Dipartimento di Fisica “E.R. Caianiello”, Università di Salerno, Via Giovanni Paolo II 132, I-84084 Fisciano, Italy}
    \affiliation{Istituto Nazionale di Fisica Nucleare, Sezione di Napoli, Via Cintia, 80126, Napoli, Italy}
    \affiliation{Nordic Institute for Theoretical Physics (NORDITA), Hannes Alfv\'{e}ns v\"{a}g 12, SE-114 19 Stockholm, Sweden}

    \author{Fabio Aratore}
    \email[]{faratore@unisa.it}
    \affiliation{Dipartimento di Fisica “E.R. Caianiello”, Università di Salerno, Via Giovanni Paolo II 132, I-84084 Fisciano, Italy}
    \affiliation{Istituto Nazionale di Fisica Nucleare, Sezione di Napoli, Via Cintia, 80126, Napoli, Italy}

    \author{Valerio Bozza}
    \email[]{vbozza@unisa.it}
    \affiliation{Dipartimento di Fisica “E.R. Caianiello”, Università di Salerno, Via Giovanni Paolo II 132, I-84084 Fisciano, Italy}
    \affiliation{Istituto Nazionale di Fisica Nucleare, Sezione di Napoli, Via Cintia, 80126, Napoli, Italy}

\begin{abstract}
    This paper investigates gravitational lensing in the strong deflection limit, focusing particularly on higher-order images produced near compact objects such as black holes and their observable impact through the visibility function. Employing a robust parametrization framework proposed by Rezzolla and Zhidenko, the study systematically explores deviations from the Schwarzschild metric. A detailed theoretical analysis of interferometric observables is provided, highlighting how higher-order images imprint distinctive, measurable patterns in the visibility function, notably characterized by a staircase-like structure.\ By parametrically varying metric coefficients, the analysis reveals clear dependencies between spacetime deviations and key observational signatures, specifically the step heights and periodicities in the interferometric visibility. The results enhance the theoretical groundwork for interpreting data from advanced interferometric observations, potentially enabling precise tests of general relativity and the discrimination among alternative gravitational theories.
\end{abstract}


\maketitle

\section{Introduction}
\label{sec:intro}

According to the theory of general relativity, photons follow curved paths rather than straight lines when moving through curved spacetime \cite{Weinberg1972,Misner1973}.\ This phenomenon, known as gravitational light deflection, offers a unique observational and theoretical tool in astrophysics \cite{Falco1992,Narayan,Wambsganss1998,Schneider2001,Dodelson2017,Meneghetti2021}. While such deflection takes place around any gravitating mass, it is useful to distinguish between two principal regimes of analysis. When light rays pass relatively far from the object, the weak deflection approximation can be employed. This regime has proven remarkably effective in explaining a broad array of observational data, such as gravitational lensing by galaxies \cite{Walsh} and galaxy clusters \cite{Soucail}. However, when photons pass in the vicinity of a compact object such as a black hole, the deflection becomes so large that the weak-deflection expansion fails. In these cases, one must rely on the so-called strong deflection limit approximation \cite{Bozza2001}.

Historically, the strong deflection limit approximation was first introduced by Darwin for the Schwarzschild geometry \cite{Darwin1959}, showing a logarithmic divergence in the deflection angle as light approaches the photon sphere — the spherical region covered by all unstable circular photon orbits with arbitrary inclination angles. In this limit, photons may loop around the black hole before eventually escaping, generating an infinite sequence of so-called higher-order images that accumulate exponentially close to the edge of the black hole shadow \cite{Atkinson1965,Luminet1979,Ohanian1987}.

The foundational work of Darwin was later extended, leading to broader formulations applicable to generic spherically symmetric spacetimes \cite{Bozza2002}, as well as rotating spacetimes \cite{Bozza2003} (on the numerical side, it is worth mentioning the work of Virbhadra and Ellis \cite{Ellis2000}, who calculated the properties of higher-order images starting from the exact expression for the deflection angle in the case of a Schwarzschild black hole, as well as the work of Frittelli, Kling, and Newman \cite{Frittelli2000}, who obtained solutions to the exact lens equation in the form of integral expressions; see also the work of Perlick \cite{Perlick2004}). Since then, a growing number of studies have explored the strong-deflection regime, particularly in the context of testing alternative theories of gravity, see, e.g., Refs.\ \cite{Claudel2001, Hasse2002, Perlick2004-review, Iyer2007, Keeton2008, Tsupko2008, Majumdar2009, Tarasenko2010, Eiroa2011, Wei2012, Zhang2015, Alhamzawi2016, Tsukamoto2016, Aldi-Bozza-2017, Dai2018, Kuang2022, Aratore-Bozza-2024,Guo2025}. Notably, considerable attention has been given to higher-order images, see, e.g., \cite{Gralla2019, Johnson-2020, Gralla2020, Lupsasca2020, Gralla-Lupsasca-2020, Wielgus-2021, Broderick-2022, Ayzenberg-2022, Guerrero-2022, BK-Tsupko-2022, Tsupko-2022, Eichhorn-2023, Broderick-Salehi-2023,Kocherlakota-2024-2, Aratore-Tsupko-Perlick-2024}. Moreover, generalizations of the strong deflection limit analysis have incorporated more realistic astrophysical scenarios, such as the influence of cold, non-magnetized plasma, both homogeneous \cite{Tsupko2013} and inhomogeneous \cite{Feleppa2024plasma}; recently, a strong deflection limit analysis for massive particles has also been developed \cite{Feleppa2024massive}. These extensions have enriched the applicability of the procedure to wider contexts.

Observationally, the Event Horizon Telescope team recently conducted groundbreaking high-resolution campaigns that, for the first time, directly imaged the supermassive black holes M87* \cite{L1,L2,L3,L4,L5,L6} and Sgr A* \cite{L12,L13,L14,L15,L16,L17}. These observations have significantly advanced our ability to test general relativity in the strong-field regime. Furthermore, polarimetric images obtained by the Event Horizon Telescope team have further revealed the presence and morphology of magnetic fields in the near-horizon region, offering key insights into accretion dynamics and jet formation \cite{pol1,pol2}. These findings have confirmed the presence of supermassive black holes at galactic centers and initiated a new era in testing general relativity in previously inaccessible regimes. Event Horizon Telescope data have already enabled an initial estimate of the spin of M87* by analyzing the orbital angular momentum content of the observed radiation through wavefront phase reconstruction techniques \cite{Tamburini1,Tamburini2}; see also Ref.\ \cite{Tamburini3}.

Motivated by these advances, in Ref.\ \cite{Aratore2021}, the authors apply the formalism of the strong deflection limit to spherically symmetric spacetimes, computing the positions and angular separations of higher-order images and connect these quantities to observable signatures in the complex visibility function — a central interferometric quantity defined as the Fourier transform of the source's brightness distribution; it has been demonstrated that higher-order images imprint distinctive features on the visibility function, most notably a characteristic staircase pattern. Such structure encodes crucial details about the spacetime metric, thus potentially providing direct observational evidence for testing gravitational theories. This characteristic staircase structure in the complex visibility function had previously been highlighted in the literature, with pioneering contributions by Johnson and collaborators \cite{Johnson-2020}. 

Complementing these theoretical developments, Rezzolla and Zhidenko proposed a robust parametrization for spherically symmetric black hole metrics in generic metric theories of gravity \cite{Rezzolla2014}. This approach systematically captures deviations from the Schwarzschild solution and facilitates model-independent tests of general relativity. Unlike traditional expansions in inverse radial coordinates \cite{Johannsen2011,Cardoso2014}, the Rezzolla-Zhidenko metric employs a compactified radial coordinate combined with a continued-fraction expansion. One of the key advantages of this parametrization is its improved convergence properties, which reduce the number of parameters required to accurately describe the spacetime metric from the event horizon to asymptotic infinity. The Rezzolla-Zhidenko parametrization has been applied in different contexts, see, e.g., Refs.\ \cite{delaurentis2023,Ahmedov2023,Malafarina2025,Moriyama2025}.

In the present study, we integrate the results obtained in Ref.\ \cite{Aratore2021} with the parametric framework introduced by Rezzolla and Zhidenko \cite{Rezzolla2014}. Specifically, we apply such parametrization to investigate detailed properties of higher-order images and their interferometric signatures.\ Ultimately, this methodology provides a powerful theoretical framework to interpret high-resolution observational data, enhancing our ability to distinguish among competing gravitational theories through future advanced interferometric experiments.

The paper is organized as follows. In Sec.\ II, we summarize the main results on gravitational deflection in the strong deflection limit for spherically symmetric spacetimes. In Sec.\ III, we introduce the key interferometric observable: the visibility function. In Sec.\ IV, we present the continued-fraction parametrization of static, spherically symmetric metrics proposed by Rezzolla and Zhidenko \cite{Rezzolla2014}, outlining its relevance and advantages over other parametrizations proposed in the literature. Sec.\ V is devoted to applying the strong deflection limit formalism within this parametrized framework, eventually focusing on how variations in the metric parameters affect observable quantities such as the step height and periodicity in the visibility function. In Sec.\ VI, we derive and discuss perturbative expressions for all relevant quantities, providing a clearer interpretation of the parameter dependencies and their phenomenological implications. Finally, Sec.\ VII contains our concluding remarks.

\section{Gravitational lensing in the strong deflection limit}

Following Refs.\ \cite{Aratore2021,BozzaScarpetta2007}, in this section we briefly summarize the main equations of gravitational lensing by black holes in the strong deflection limit, focusing on the spherically symmetric case and thus simplifying the analysis of interferometric observables in the subsequent sections.

In a spherical polar coordinate system $(t, r, \vartheta, \phi)$, a general static and spherically symmetric black-hole metric can be expressed as
\begin{equation} \label{line element ssbhs}
    \mathrm{d}s^{2} = -A(r)dt^{2} + B(r) \mathrm{d}r^{2} + C(r)\mathrm{d}\Omega^2 \, ,
\end{equation}
with $C(r) = r^2$ and where $\mathrm{d}\Omega^2 \coloneqq \mathrm{d}\vartheta^{2} + \sin^{2}\vartheta \, \mathrm{d}\varphi^{2}$ defines the round metric on the unit two-sphere. We also assume the spacetime to be asymptotically flat (we neglect any cosmological effect), so the metric coefficients $A(r)$ and $B(r)$ satisfy
\begin{equation}
    \lim_{r \to \infty} A(r) ~ = ~ 1 \, , \quad
    \lim_{r \to \infty} B(r) ~ = ~ 1 \, .
\end{equation}
Without loss of generality, motion is restricted to the equatorial plane by setting $\vartheta = \pi/2$. Additionally, we further assume that the equation
\begin{equation} \label{photon sphere equation}
    \frac{C^\prime(r)}{C(r)} = \frac{A^\prime(r)}{A(r)}
\end{equation}
admits at least one positive solution. The largest root of this equation represents the radius of the photon sphere and will be denoted by $r_m$.

Since we are interested in photon trajectories with a closest approach distance $r_0$ near $r_m$, we introduce a parameter $\delta \ll 1$ such that
\begin{equation}
    r_0 \coloneqq r_m (1 + \delta) \, .
\end{equation}
The corresponding impact parameter $u$ of the photon, related to $r_0$ by the equation
\begin{equation} \label{impact parameter}
    u = \sqrt{\frac{C(r_0)}{A(r_0)}} \, ,
\end{equation}
must also be close to its minimum value $u_m$. Therefore, to express this deviation, we define a parameter $\varepsilon \ll 1$ through the relation
\begin{equation}
    u \coloneqq u_m (1 + \varepsilon).
\end{equation}
The two parameters $\varepsilon$ and $\delta$ are related by
\begin{equation}
    \varepsilon \simeq \frac{\beta_m}{2 u_m^2} \delta^2 \, ,
\end{equation}
with the coefficient $\beta_m$ given by
\begin{equation}
    \beta_m = \frac{r_m^2}{2} \frac{C_m^{\prime \prime} A_m - A_m^{\prime \prime} C_m}{A_m^2} \, . 
\end{equation}
In the above equation, the prime indicates differentiation with respect to the radial coordinate, while the subscript $m$ signifies that the result should be evaluated at $r = r_m$.

Consider now a source at a radial coordinate $r_S$ and an observer at $r_O$. Up to first order in $\epsilon$, the azimuthal shift experienced by a photon along its path is given by
\begin{equation} \label{azimuthal shift}
    \Delta \varphi = -\Bar{a} \log \frac{\varepsilon}{\eta_O \eta_S} + \Bar{b} \, ,
\end{equation}
where the quantities $\eta_O$ and $\eta_S$ are defined as
\begin{equation}
    \eta_O \coloneqq 1 - \frac{r_m}{r_O} \, , \quad \eta_S \coloneqq 1 - \frac{r_m}{r_S} \, ,
\end{equation}
while the strong deflection limit coefficients $\Bar{a}$ and $\Bar{b}$ read
\begin{gather}
    \Bar{a} \coloneqq r_m \sqrt{\frac{B_m}{A_m \beta_m}} \, , \label{SDLcoeffa} \\
    \Bar{b} \coloneqq \Bar{a} \log\left(\frac{2 \beta_m}{u_m^2}\right) + \Bar{b}_O + \Bar{b}_S \, , \label{coefficient b} \\
    \Bar{b}_O \coloneqq \int_0^{\eta_O} g_1(\eta) d\eta \, , \\
    \Bar{b}_S \coloneqq \int_0^{\eta_S} g_1(\eta) d\eta \, ,
\end{gather}
where, in turn, the function $g_1(\eta)$ is expressed as
\begin{multline}
    g_1(\eta) \coloneqq u_m \sqrt{\frac{B(r(\eta))}{C(r(\eta))}} \left( \frac{C(r(\eta))}{A(r(\eta))} - 1 \right)^{-1/2} \frac{r_m}{(1 - \eta)^2} \\
    - \frac{u_m}{\sqrt{\beta_m}} \sqrt{\frac{B_m}{C_m}} \frac{r_m}{\eta} \, . 
\end{multline}
Denoting by $\varphi_S \in [-\pi, \pi]$ the azimuthal coordinate of the source and conveniently fixing the origin of the coordinate system such that $\varphi_O = \pi$, with $\varphi_O$ being the azimuthal coordinate of the observer, we find from the lens equation for spherically symmetric compact objects,
\begin{equation}
    \varphi_O - \varphi_S = \Delta \varphi + 2 n \pi \, ,
\end{equation}
as well as from Eq.\ \eqref{azimuthal shift}, the position of the images as
\begin{equation} \label{positionsimages}
    \varepsilon_{n,\pm} = \eta_O \eta_S \exp{\left[\frac{\Bar{b} \pm \varphi_S - (2n + 1)\pi}{\Bar{a}}\right]} \, ,
\end{equation}
where $n$ represents the number of full revolutions completed by the photons before reaching the observer, while the double sign reflects whether the images form on the same side as the source (positive parity) or on the opposite side (negative parity). We remind the reader that the strong deflection limit becomes exact as $n$ approaches infinity; however, it typically serves as a very accurate approximation even for $n = 1$.

For an asymptotic observer, the angle between the photon’s path and the line pointing toward the black hole is given by $\theta = u/r_O$; thus, from Eq.\ \eqref{positionsimages}, we obtain
\begin{equation} \label{thetam}
    \theta_{n,\pm} = \theta_m \left( 1 + \varepsilon_{n,\pm} \right) \, ,
\end{equation}
where $\theta_m = u_m /r_O$ is usually referred to as the angular radius of the shadow of the black hole (strictly speaking, such terminology is not entirely accurate:~gravitomagnetic simulations favor an emitting region below the photon sphere \cite{Nazarova2019,Nazarova20201,Nazarova20202,Chael2021}; however, for the sake of simplicity, we will continue referring to $\theta_m$ as the radius of the shadow). As can be seen from the above equation, higher-order images move exponentially closer to $\theta_m$ as the number of loops increases.

It is also worth noting that Eqs.\ \eqref{azimuthal shift}, \eqref{positionsimages} and \eqref{thetam} remain valid even for sources located within the photon sphere.

In the following section, we will introduce the main actor in interferometric observations: the visibility function.

\section{Complex visibility function of a compact source} \label{vfcompactsource}

High-resolution interferometric observations measure the visibility function, which is the Fourier transform of the source’s surface brightness distribution $I(x, y)$, with $x$ and $y$ being the angular coordinates in the observer’s sky. If we consider a static source with a Gaussian profile, then the images are distributed along a single axis, say the $x$-axis. The visibility function, expressed in terms of the spatial frequencies $(u, v)$\footnote{The components $u$ and $v$ represent the interferometric baseline projected perpendicular to the line of sight, measured in units of the observing wavelength $\lambda$.}, Fourier conjugates of the angular coordinates $(x, y)$, can then be written as \cite{Aratore2021}
\begin{multline}
    V(u, v) = N(v) \sum_{p = \pm} \sum_{n = 1}^{\infty} \varepsilon_{n, p} \, \exp\left({-2 \pi^2 \Delta \theta_0^2 \, \varepsilon_{n,p}^2 u^2}\right) \\
    \times \exp\left({-2 \pi i p \, \theta_{n, p} u}\right) \, ,
\label{visibility function}
\end{multline}
where we defined
\begin{align}
    N(v) &\coloneqq 2 \pi I_0 \theta_m \Delta \theta_0 \Delta \vartheta_S e^{-2 \pi^2 \theta_m^2 \Delta \vartheta_S^2 \, v^2} \, , \\
    \Delta \theta_0 &\coloneqq \frac{\theta_m}{\bar a} \Delta \varphi_S + \frac{\theta_m r_m}{r_S^2} \left( \frac{1}{\eta_S} + \frac{g_1(\eta_S)}{\bar a} \right) \Delta r_S \, .
\end{align}
In the above expressions, $I_0$ is the central brightness of the source, while $\Delta \varphi_S$, $\Delta \vartheta_S$ and $\Delta r_S$ represent its azimuthal size, polar size and radial extension, respectively. As we can see, the $v$-dependence of the visibility function is contained in the quantity $N(v)$, which does not depend on $n$ and is the same for all contributions\footnote{Note that, under the assumption of spherical symmetry, the tangential extension of the images is the same as that of the source, allowing it to be factored out of the sum.}. The remaining quantities appearing in the visibility are $n$-dependent and encapsulate all the details of the higher-order images, which, in turn, are characteristic of the black-hole metric.

The visibility function of a compact source around a black hole exhibits a staircase structure, where each step is determined by a single term in the sum in Eq.\ \eqref{visibility function}. We identify two key pieces of information from the visibility function. The first is the step height, which corresponds to the amplitude in Eq.\ \eqref{visibility function}, expressed as \cite{Aratore2021}
\begin{equation} \label{step height} 
h_{n, \pm} = N(v) \, \varepsilon_{n, \pm} \, . 
\end{equation}
The second is encoded in the phase factor. The interference between images on opposite sides of the black hole — separated by a distance $\theta_{n, +} + \theta_{n, -}$ for images of the same order, or $\theta_{n, -} + \theta_{n+1, +}$ for adjacent orders — induces modulations in the visibility amplitude. These modulations have characteristic periodicities given by \cite{Aratore2021}
\begin{align} 
P_{n,+} &= \left[ \theta_m \left( 2 + \varepsilon_{n, +} + \varepsilon_{n, -} \right) \right]^{-1} \, , \label{psameorder} \\ 
P_{n,-} &= \left[ \theta_m \left( 2 + \varepsilon_{n, -} + \varepsilon_{n+1, +} \right) \right]^{-1} \, .
\end{align}
These characteristic features of the visibility function can thus be used to extract the properties of the black hole metric.

With the aim of analyzing how potential deviations from general relativity could show up in the interferometric signature of compact objects, we proceed in the next section to briefly review the parametrization of spherically symmetric spacetimes introduced in Ref.\ \cite{Rezzolla2014}.

\section{Parametrization of spherically symmetric black holes} \label{parametrizationSSBHs}

We review here the continued-fraction parametrization framework introduced in Ref.\ \cite{Rezzolla2014}.\ The Rezzolla-Zhidenko metric reproduces static, spherically symmetric and asymptotically flat black hole solutions with an event horizon, in both general relativity and generic alternative metric theories of gravity. This method provides an accurate representation of the metric functions across the entire spacetime, from the horizon to asymptotic infinity. 

For the purpose of this section, we express the coefficient $B(r)$ in Eq.\ \eqref{line element ssbhs} as $B(r) = N(r)/A(r)$. As a consequence, the line element is rewritten as
\begin{equation} \label{line element rewritten}
    \mathrm{d}s^{2} = -A(r)dt^{2} + \frac{N(r)}{A(r)}\mathrm{d}r^{2} + r^{2}\mathrm{d}\Omega^2 \, .
\end{equation}

The horizon of the black hole is located at the largest positive root of the equation $A(r) = 0$; its radius will hereafter be denoted as $r_{H}$. To achieve better convergence properties of the approximation method, the radial coordinate $r$ is compactified by introducing the variable
\begin{equation}
    x(r) \coloneqq 1 - \frac{r_H}{r},
\end{equation}
which smoothly maps the horizon $(r = r_{H})$ to $x = 0$ and spatial infinity $(r \to \infty)$ to $x = 1$. In terms of this variable, the metric coefficients may be rewritten as
\begin{equation}
    A(r) = x \mathcal{A}(x) \, , \quad
    N(r) = \mathcal{B}(x)^2 \, ,
\end{equation}
with the functions $\mathcal{A}(x)$ and $\mathcal{B}(x)$, strictly positive in the range $0 \le x \le 1$, defined as
\begin{align}
    \mathcal{A}(x) &\coloneqq 1 - \epsilon (1 - x) + (a_0 - \epsilon)(1 - x)^2 \nonumber \\
    &\hspace{3.5cm} + \Tilde{\mathcal{A}}(x) (1 - x)^3, \label{A(x)} \\
    \mathcal{B}(x) &\coloneqq 1 + b_0 (1 - x) + \Tilde{\mathcal{B}}(x) (1 - x)^2 \, , \label{B(x)}
\end{align}
where, in turn, the functions $\Tilde{\mathcal{A}}(x)$ and $\Tilde{\mathcal{B}}(x)$ are expanded in terms of continued fractions:
\begin{align} \label{ABtilde}
    \Tilde{\mathcal{A}}(x) &= \dfrac{a_1}{1 + \dfrac{a_2 x}{1 + \dfrac{a_3 x}{1 + \ldots}}} \, , \\
    \Tilde{\mathcal{B}}(x) &= \dfrac{b_1}{1 + \dfrac{b_2 x}{1 + \dfrac{b_3 x}{1 + \ldots}}} \, .
\end{align}
The parameters appearing in the above expressions, namely Eqs.~\eqref{A(x)}--\eqref{ABtilde}, can be classified into two distinct categories.\ Apart from the parameter $\epsilon > -1$, which quantifies deviations of the horizon radius from the Schwarzschild radius, $\epsilon = -1 + 2M/r_H$, the first set, consisting of $a_0$ and $b_0$, is determined by the asymptotic behavior of the metric at spatial infinity. The second set, involving the continued-fraction coefficients $(a_1, a_2, \ldots, b_1, b_2, \ldots)$, is constrained by considering phenomena happening near the event horizon.

In the following, we take into account only the first few leading-order parameters, i.e., the metric will depend only on $\epsilon, a_0, b_0, a_1$ and $b_1$. While the inclusion of additional expansion coefficients does not extend the class of metrics represented by spherically symmetric, asymptotically flat black hole solutions with an event horizon, it would certainly improve the accuracy with which diverse solutions within this class can be represented. In terms of the coordinate $r$, the coefficients $A(r)$ and $N(r)$ are then given by \cite{Rezzolla2024}
\begin{align}
   A(r) &= 1 - \frac{2M}{r} + \frac{4 a_0}{(1 + \epsilon)^2}\frac{M^2}{r^2} \nonumber \\
   &\hspace{1cm} + \frac{8(\epsilon - a_0 + a_1)}{(1 + \epsilon)^3}\frac{M^3}{r^3} - \frac{16 a_1}{(1 + \epsilon)^4}\frac{M^4}{r^4} \, , \label{coeffA} \\
   N(r) &= \left[1 + \frac{2 b_0}{1 + \epsilon}\frac{M}{r} + \frac{4 b_1}{(1 + \epsilon)^2}\frac{M^2}{r^2}\right]^2 \, . \label{coeffN}
\end{align}

For later convenience, let us now consider two specific examples. The first is the Reissner-Nordström black hole, described by the line element \cite{Reissner1916,Nordström1918}
\begin{multline}
    \mathrm{d}s^{2} = -\left( 1 - \frac{2 M}{r} + \frac{q^2}{r^2} \right) dt^{2} 
    \\
    + \left( 1 - \frac{2 M}{r} + \frac{q^2}{r^2} \right)^{-1} \mathrm{d}r^{2} + r^{2} \mathrm{d}\Omega^{2} \, ,
\end{multline}
where $q$ is the charge parameter. The coefficients in the parametrized framework that reproduce the above metric are given by
\begin{align} \label{parRN}
    b_0 &= a_1 = b_1 = 0 \, , \\
    a_0 &= \epsilon = \frac{2}{1 + \sqrt{1 - q^2}} - 1 \, .
\end{align}
The second example is the dilaton black hole (in the absence of both the axion field and spin), described by the line element \cite{García1995}
\begin{multline} \label{metric dilaton BH}
    \mathrm{d}s^{2} ~ = ~ -\left( \frac{\rho - 2\mu}{\rho + 2b_{\text{dil}}} \right) dt^{2} 
    + \left( \frac{\rho + 2 b_{\text{dil}}}{\rho - 2\mu} \right) \mathrm{d}r^{2} 
    \\
    + (\rho^2 + 2 b_{\text{dil}} \, \rho)\mathrm{d}\Omega^{2} \, ,
\end{multline}
\begin{figure*}[!t]
\includegraphics[width=9cm]{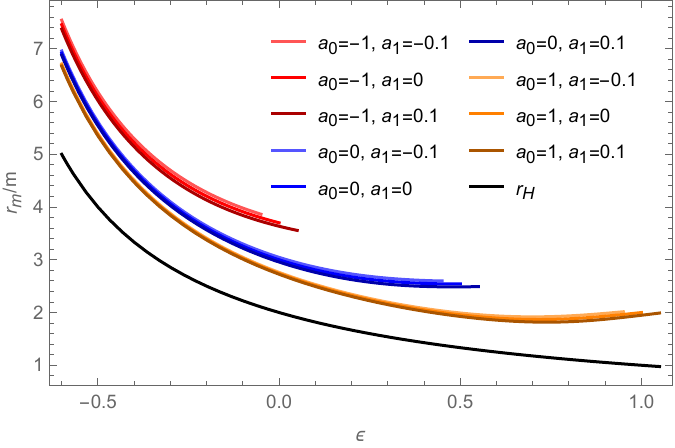}
\centering
\caption{Behavior of the photon sphere radius $r_m$ as a function of $\epsilon$, across different values of $a_0$ and $a_1$.}
\end{figure*}

\begin{figure*}[!t]
\includegraphics[width=9cm]{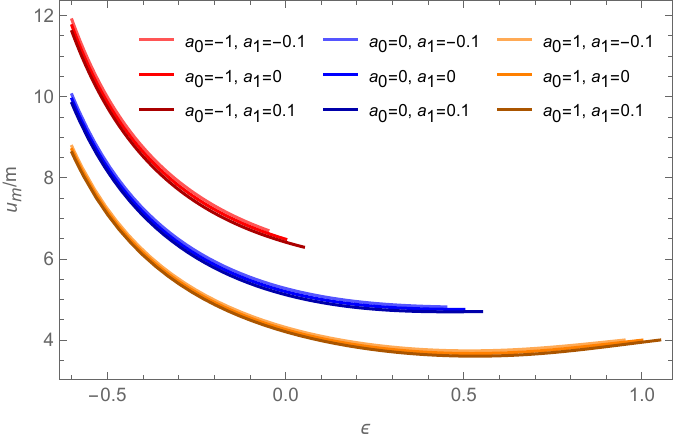}
\centering
\caption{Behavior of the minimum impact parameter $u_m$ as a function of $\epsilon$, across different values of $a_0$ and $a_1$.}
\end{figure*}
where $b_{\text{dil}}$ is the dilaton parameter, and $\mu = M - b_{\text{dil}}$, with $M$ being the ADM mass. The features of such a spacetime can be reproduced by the following expressions for the coefficients $a_0$, $b_0$, $a_1$ and $b_1$ in the parametrized scheme:
\begin{align}
    \epsilon &= \sqrt{1 + \frac{b_{\text{dil}}}{\mu}} - 1 \, , \label{epsdil} \\
    a_0 &= \frac{b_{\text{dil}}}{2\mu} \, , \\
    b_0 &= 0 \, , \\
    a_1 &= 2 \sqrt{1 + \frac{b_{\text{dil}}}{2\mu}} + \frac{1}{1 + \dfrac{b_{\text{dil}}}{2\mu}} - 3 - \dfrac{b_{\text{dil}}}{2\mu} \, , \\
    b_1 &= \dfrac{1 + \dfrac{b_{\text{dil}}}{\mu}}{1 + \dfrac{b_{\text{dil}}}{2\mu}} - 1 \, . \label{b1dil}
\end{align}
These metrics provide notable examples of spacetimes that can be reproduced by the parametrization.

\subsection{Theoretical constraints}

As pointed out in Refs.\ \cite{Rezzolla2014,Rezzolla2020}, non-trivial theoretical constraints on the allowed ranges of the parameters $\epsilon$, $a_i$ and $b_i$ must be carefully taken into account. In the next section, where we plot various quantities of interest as functions of $\epsilon$ for different values of $a_i$ and/or $b_i$, the truncation of each curve directly reflects these theoretical constraints. To be more specific, the metric function $N(r)^2$ — which defines the lapse in the line element of a static, spherically symmetric spacetime — must admit a single, non-degenerate, outermost Killing horizon. This requires that the equation $N(r)^2 = 0$ possesses a largest real, positive root $r_0$, and that $N(r)^2 > 0$ for all $r > r_0$. To identify the domain of $\epsilon$ for which these conditions are satisfied, we solve the equation $N(r)^2 = 0$ and compute all real roots as functions of $\epsilon$, for fixed values of the parameters $a_0, a_1, b_0$ and $b_1$. For each value of $\epsilon$, we identify the largest real positive root as the horizon radius. As $\epsilon$ varies, additional real positive roots may appear. We follow the evolution of all real roots and determine the value of $\epsilon$ at which the root we have been tracking intersects another, indicating that it no longer represents the original outermost horizon. In this way, we systematically delimit the allowed parameter space shown in the plots. Moreover, as already pointed out in Ref.\ \cite{Rezzolla2022}, one also has to ensure that the additional metric function $B(r)^2$, which depends on the $b_i$ coefficients, remains nonvanishing for all $r \geq r_0$, thereby preventing divergences in the curvature invariants and maintaining the regularity of the spacetime.

\section{Strong deflection limit analysis in a parametrized scheme}

\begin{figure}[!t]
\includegraphics[width=8.65cm]{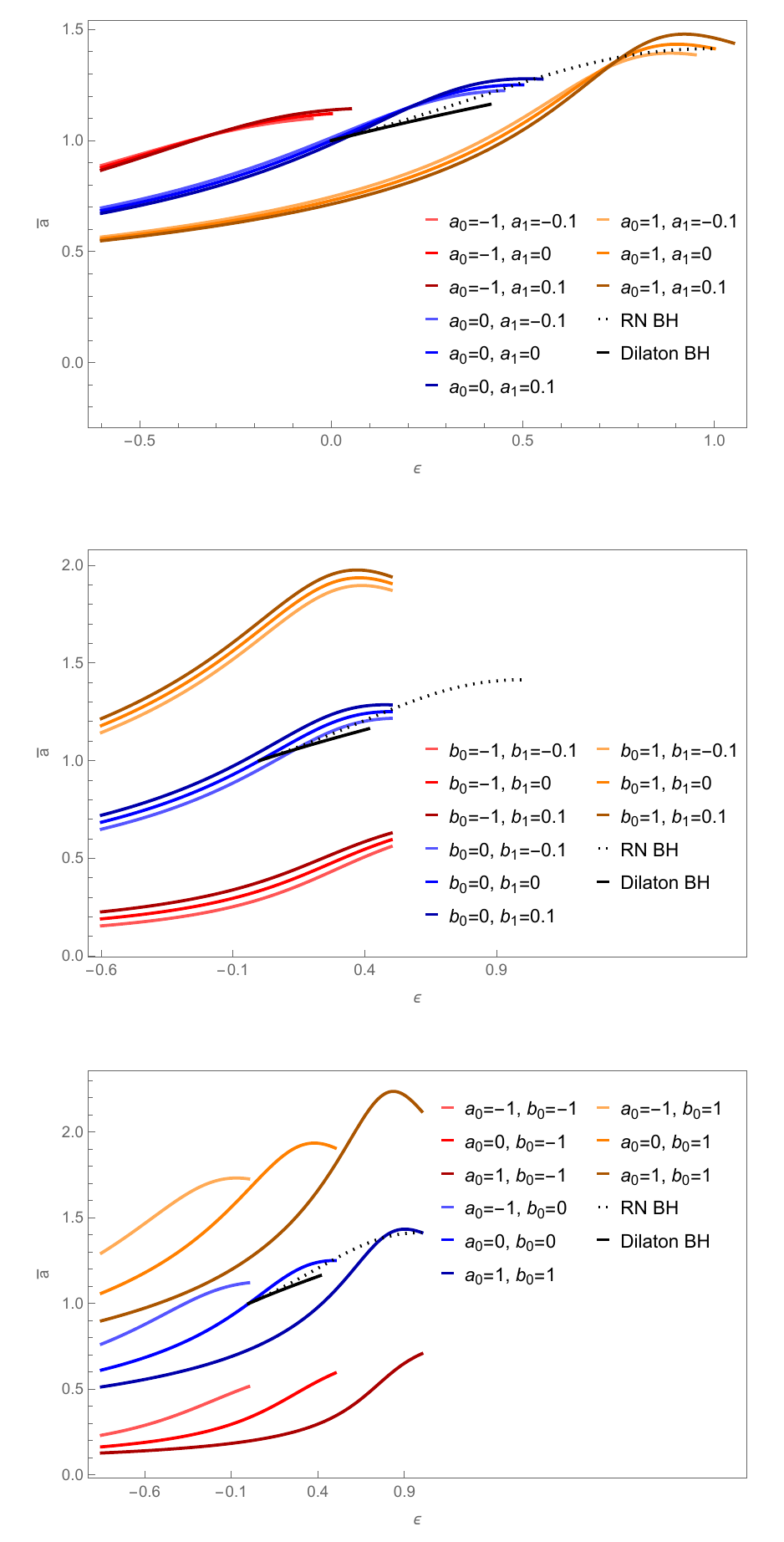}
\centering
\caption{Behavior of the strong deflection limit coefficient $\Bar{a}$ as a function of $\epsilon$. Different families of curves correspond to different choices of the parameters $a_0$, $a_1$, $b_0$ and $b_1$, as indicated in the legends. The black solid line represents the dilaton black hole, while the black dashed line corresponds to the Reissner-Nordström black hole. In the legend, ``RN BH'' stands for ``Reissner-Nordström black hole''.}
\end{figure}

\begin{figure}[!t]
\includegraphics[width=8.65cm]{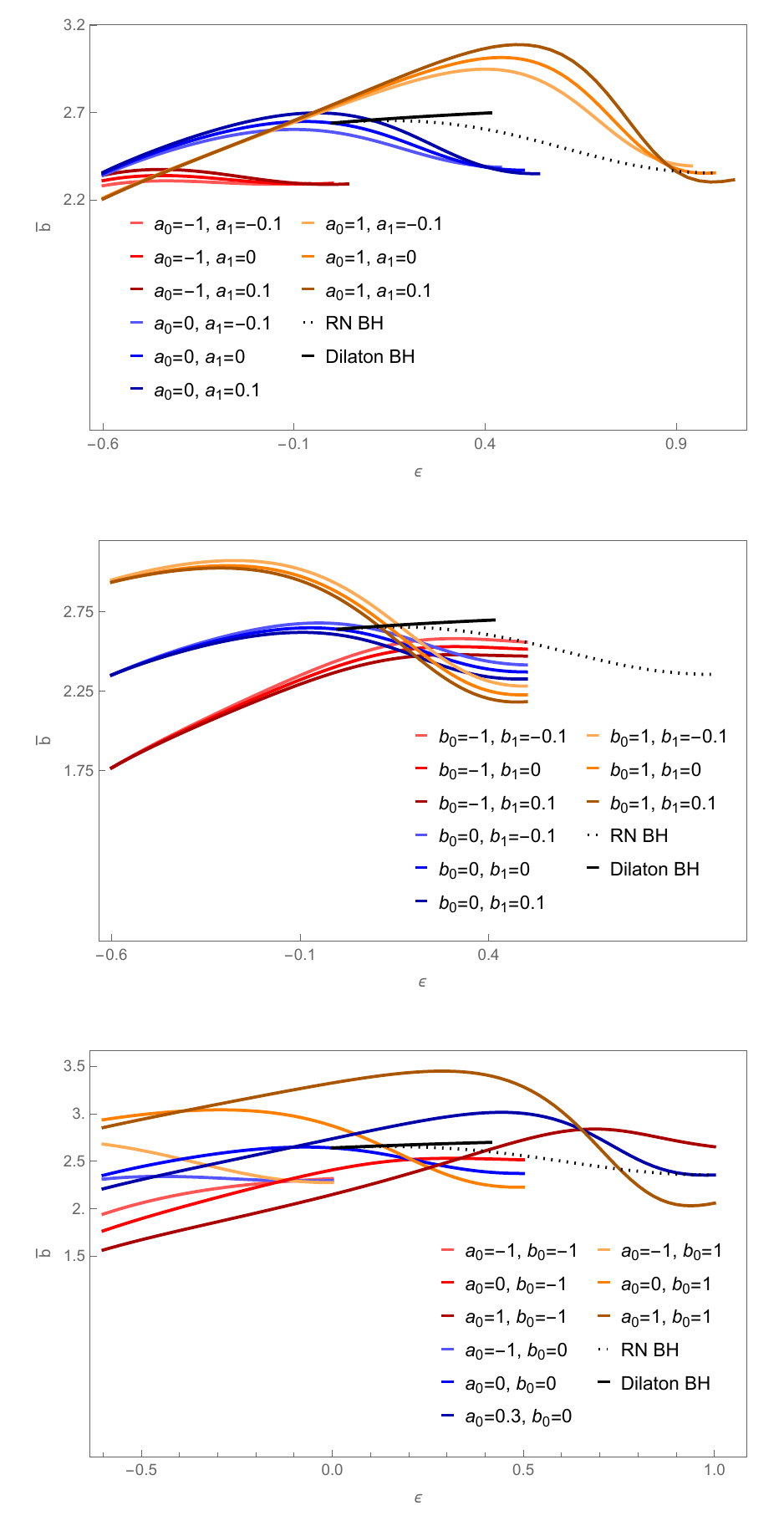}
\centering
\caption{Behavior of the strong deflection coefficient $\Bar{b}$ as a function of $\epsilon$, across different values of the parameters $a_0$, $a_1$, $b_0$ and $b_1$. The black solid line represents the dilaton black hole, while the black dashed line corresponds to the Reissner-Nordström black hole. We set $\eta_O = 1$ and $r_S$ to $20M$. In the legend, ``RN BH'' stands for ``Reissner-Nordström black hole''.}
\end{figure}

In this section, starting from the metric \eqref{line element rewritten}, with $A(r)$ and $N(r)$ given by Eqs.\ \eqref{coeffA} and \eqref{coeffN}, respectively, we present the results of the strong deflection limit analysis, as well as the behavior of the step height and characteristic periodicities in the visibility function. In particular, we analyze how the interferometric signature is affected by variations in the parameters $\epsilon, a_0, b_0, a_1$ and $b_1$.

We begin by calculating the radius of the photon sphere from Eq.\ \eqref{photon sphere equation} and then use Eq.\ \eqref{impact parameter} to find the minimum impact parameter. In Figs.\ 1 and 2, we show the behavior of $r_m$ and $u_m$, respectively, as a function of $\epsilon$, for different values of $a_0$ and $a_1$ (note that both $r_m$ and $u_m$ do not depend on $b_0$ and $b_1$). Regarding the orange and blue curves, the figures show that both the radius of the photon sphere ($r_m$) and the impact parameter ($u_m$) decrease up to a certain value of $\epsilon$, beyond which they begin to rise slightly. The curves associated with a negative value of $a_0$ (red curves) instead always show a decreasing trend. In the plot for $r_m$, the radius of the event horizon has also been plotted. Regarding the other parameters, in both cases, as $a_0$ and $a_1$ increase, both the radius of the photon sphere and the minimum impact parameter decrease.

We move on by presenting, in Figs.\ 3 and 4, the results for the strong deflection limit coefficients $\Bar{a}$ and $\Bar{b}$, respectively. These quantities are plotted as functions of $\epsilon$ for different values of $a_0$, $b_0$, $a_1$, and $b_1$. Specifically, three different families have been considered: in the first one, $a_0$ and $a_1$ have been varied, in the second one, only $b_0$ and $b_1$, while in the third one, $a_0$ and $b_0$. We also note that in all the plots, the curves associated with the two examples mentioned in the previous section (dilaton and Reissner-Nordström spacetimes) have been superimposed. These curves were obtained simply by expressing the charge parameter $q$ and the dilaton parameter $b_{\text{dil}}$ in terms of $\epsilon$ (see the last expression in Eq.\ \eqref{parRN}, as well as Eq.\ \eqref{epsdil}, respectively) and then substituting the resulting formulas into the expressions for $a_0$, $b_0$, $a_1$, and $b_1$. Thus, all coefficients vary simultaneously with $\epsilon$. It is therefore interesting to compare the individual effects of variations in each coefficient (colorful curves) with the results of the coordinated variations that characterize these two physically motivated examples.

As for the lensing coefficient $\Bar{a}$, the orange and red curves in Fig.\ 3 display distinct behaviors: while the red curves increase monotonically with $\epsilon$, the orange curves show a rise followed by a drop across all three plots. The blue curves exhibit a more nuanced pattern:~in the first two plots (starting from above), $\Bar{a}$ increases and then stabilizes toward a nearly constant value, whereas in the third plot, $\Bar{a}$ rises within a certain range of $\epsilon$ and subsequently decreases. Furthermore, referring to the first plot, we observe that for fixed  $a_0$, increasing $a_1$ generally leads to an increase in the coefficient $\Bar{a}$, especially evident for 
$a_0 = 0$ and $a_1 = 0.1$. The case $a_0 = -1$ shows a relatively milder dependence on $a_1$. In the second plot, for fixed $b_0$, increasing $b_1$ consistently leads to a higher value of $\Bar{a}$, indicating a systematic trend. Similarly, in the third plot, when varying $a_0$ for fixed $b_0$, or vice versa, the coefficient 
$\Bar{a}$ also increases with both parameters, though the sensitivity to $b_0$	 appears stronger than to $a_0$.

The behavior of the strong deflection limit coefficient $\Bar{b}$ is illustrated across the three plots in Fig.\ 4, corresponding to the same families of solutions considered previously. Unlike the coefficient $\Bar{a}$, the trend of $\Bar{b}$ is significantly more intricate. This complexity arises from the fact that $\Bar{b}$ is computed through an integral over all radii, being therefore sensitive to the overall geometry, from the photon sphere to the asymptotic region.

\begin{figure}[!t]
\includegraphics[width=8.65cm]{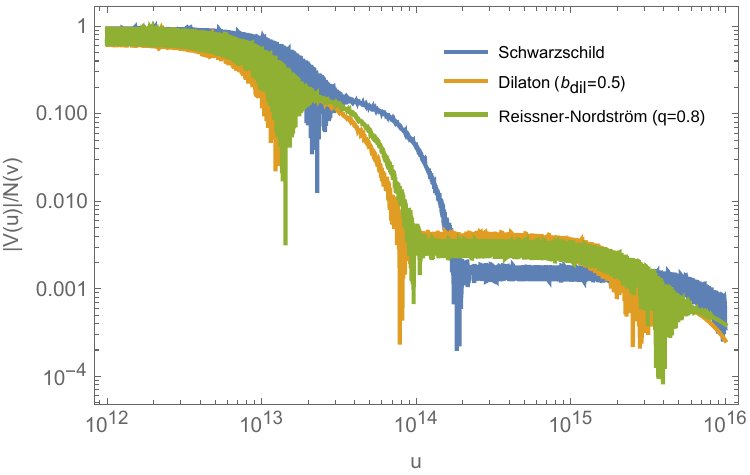}
\centering
\caption{Complex visibility function arising from higher-order images of order $n = 1, 2$, formed by lensing in the Schwarzschild, dilaton, and Reissner-Nordström spacetimes. As for the various parameters involved, they have been chosen as follows: $r_O = 16.8 \, \text{Mpc}$, $r_S = 20M$, $\Delta r_S = M$, $\varphi_S = 45^{\circ}$, $\Delta \varphi_S = \Delta r_S / r_S$.}
\end{figure}
In the first plot (again starting from above), where the parameters $a_0$ and $a_1$ vary, the red curves ($a_0 = -1$) show minimal variation with $\epsilon$, remaining nearly constant. The blue curves ($a_0 = 0$) initially increase and then decrease, while the orange curves ($a_0 = 1$) exhibit a pronounced non-monotonic behavior with a peak followed by a steep decline. Increasing $a_1$ generally raises the peak value of $\Bar{b}$, especially for higher $a_0$. In the second plot, where $b_0$ and $b_1$ vary, the curves exhibit a systematic yet nontrivial structure. For fixed $b_0$, increasing $b_1$ tends to increase the coefficient $\Bar{b}$ at lower values of $\epsilon$, but this trend reverses at larger $\epsilon$, where the curves cross and higher $b_1$ eventually results in a lower value of $\Bar{b}$. In the last plot, the interplay between $a_0$ and $b_0$ reveals complex dependencies. For instance, increasing the value of the coefficient $a_0$ while holding $b_0 = 0$ fixed (blue curves) causes $\Bar{b}$ to increase overall. In a similar way, changing $b_0$ with fixed $a_0$ (red and orange curves) leads to highly non-monotonic trends. 

\subsection{Interferometric signature}

Now that we have fully characterized lensing within a parametrized framework for spherically symmetric black holes in the strong deflection limit, we proceed to analyze the interferometric signature. To begin, in Fig.\ 5, we present the behavior of the visibility function arising from higher-order images formed by lensing in the Schwarzschild case, as well as in the two previously mentioned examples, Reissner-Nordström and dilaton. As can be seen, and as previously anticipated, the visibility function has a staircase shape; starting from the left, the first step is associated with the image corresponding to $n = 1$ (first higher-order image) with positive parity, the second with the image of the same order but with negative parity, the third with the image corresponding to $n = 2$ with positive parity, and finally the last one with the image corresponding to $n = 2$ with negative parity. The visibility function in the three cases has been normalized to its value corresponding to the image $n = 1$ with positive parity. This aims to highlight that, even if the interferometric response associated with images corresponding to $n = 1$ with positive parity were almost identical for the various metrics in question, the analysis of higher-order images would still allow distinguishing these metrics.

We proceed by showing, in Fig.\ 6, the behavior of the step heights associated with images of positive parity as a function of the parameter $\epsilon$; in particular, the three plots in the figure refer to the three families already introduced earlier, see Figs.\ 3 and 4. Starting from above, in the first plot, where $a_0$ and $a_1$ are varied, a pattern emerges that closely mirrors the trends observed for the strong deflection coefficient $\Bar{a}$. Specifically, for fixed $a_0$, increasing $a_1$ leads to an enhancement in $h_{1,+}$, with the effect being most pronounced for $a_0 = 1$ (orange curves). For $a_0 = -1$ (red curves), the values of $h_{1,+}$ remain lower and relatively flat, while for $a_0 = 0$ (blue curves), there is a noticeable growth followed by a mild plateau. In the second plot, which explores the impact of varying $b_0$ and $b_1$, a similar structure appears. For fixed $b_0$, increasing $b_1$ results in larger values of $h_{1,+}$, with the orange curves ($b_0 = 0.3$) consistently lying above the red and blue ones. In the last plot, where $a_0$ and $b_0$ vary, the interplay becomes more intricate. Here, both parameters contribute significantly to the amplitude of $h_{1,+}$, with higher values generally associated with increasing $a_0$ and $b_0$. Finally, the orange and brown curves, corresponding to $b_0 = 1$, exhibit particularly sharp peaks.

We conclude this section by presenting in Fig.\ 7 the behavior of the periodicity $P_{1,+}$ associated with the interference between images of order $n = 1$ on opposite sides of the black hole, as a function of the deformation parameter $\epsilon$. This quantity governs the frequency of modulation in the visibility function and is theoretically determined by Eq.\ \eqref{psameorder}, which depends primarily on the shadow angular radius $\theta_m = u_m / r_O$.

Across all three plots, the periodicity $P_{1,+}$ exhibits a non-monotonic trend: it increases with $\epsilon$ up to a turning point, after which it gradually decreases and tends to stabilize. This behavior is closely tied to the evolution of $\theta_m$. In the first plot starting from above, where $a_0$ and $a_1$ are varied, a clear pattern emerges: increasing either parameter leads to higher values of $P_{1,+}$, with the orange curves ($a_0 = 1$) reaching the highest values overall. In the second plot, where $b_0$ and $b_1$ vary, the influence of these parameters appears to be more limited. The transition from $b_0 = -1$ to $b_0 = 0$ produces virtually no change in the periodicity curves, and even increasing $b_1$ induces only modest shifts, suggesting that the parameters $b_0$ and $b_1$ have a subdominant effect on $P_{1,+}$. In the last plot, where both $a_0$ and $b_0$ vary, the dominant role of $a_0$ is again evident. For fixed $b_0$, increasing $a_0$ results in larger periodicities. Although increasing $b_0$ also causes some uplift in $P_{1,+}$, the variation is comparatively smaller than that driven by $a_0$.

\begin{figure}[!t]
\includegraphics[width=8.65cm]{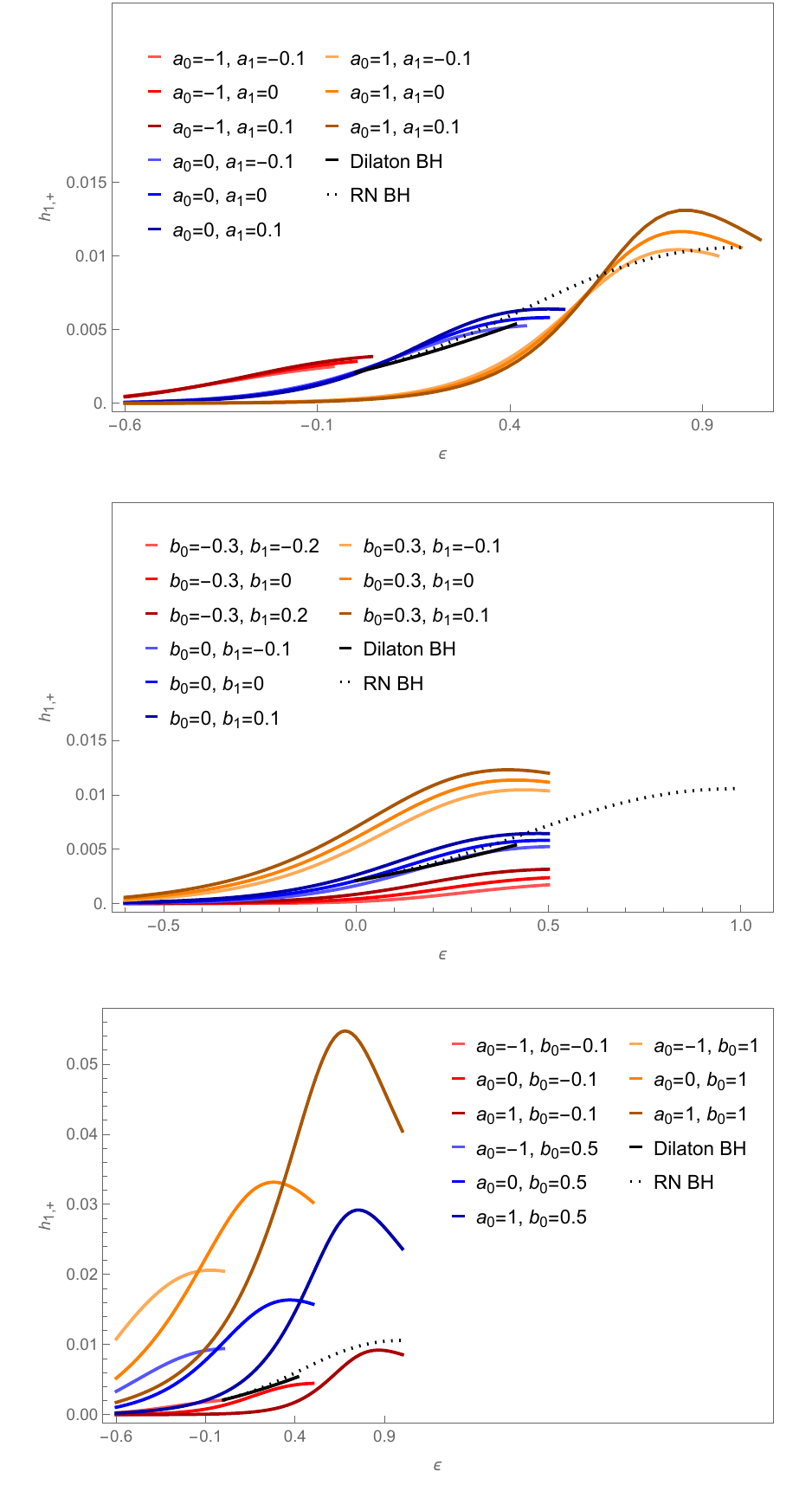}
\centering
\caption{Height of the step for the images of order $n = 1$ as a function of $\epsilon$. Different families of curves correspond to different choices of the parameters $a_0$, $a_1$, $b_0$ and $b_1$, as indicated in the legends. The black solid line represents the height of the step for the dilaton black hole, while the black dashed line corresponds to the Reissner-Nordström black hole. We set the source's azimuthal position $\varphi_S$ to $45^{\circ}$, $\eta_O = 1$ and $r_S$ to $20M$. Moreover, we also set $N(v) = 1$. In the legend, ``RN BH'' stands for ``Reissner-Nordström black hole''.}
\end{figure}

\begin{figure}[!t]
\includegraphics[width=8.6cm]{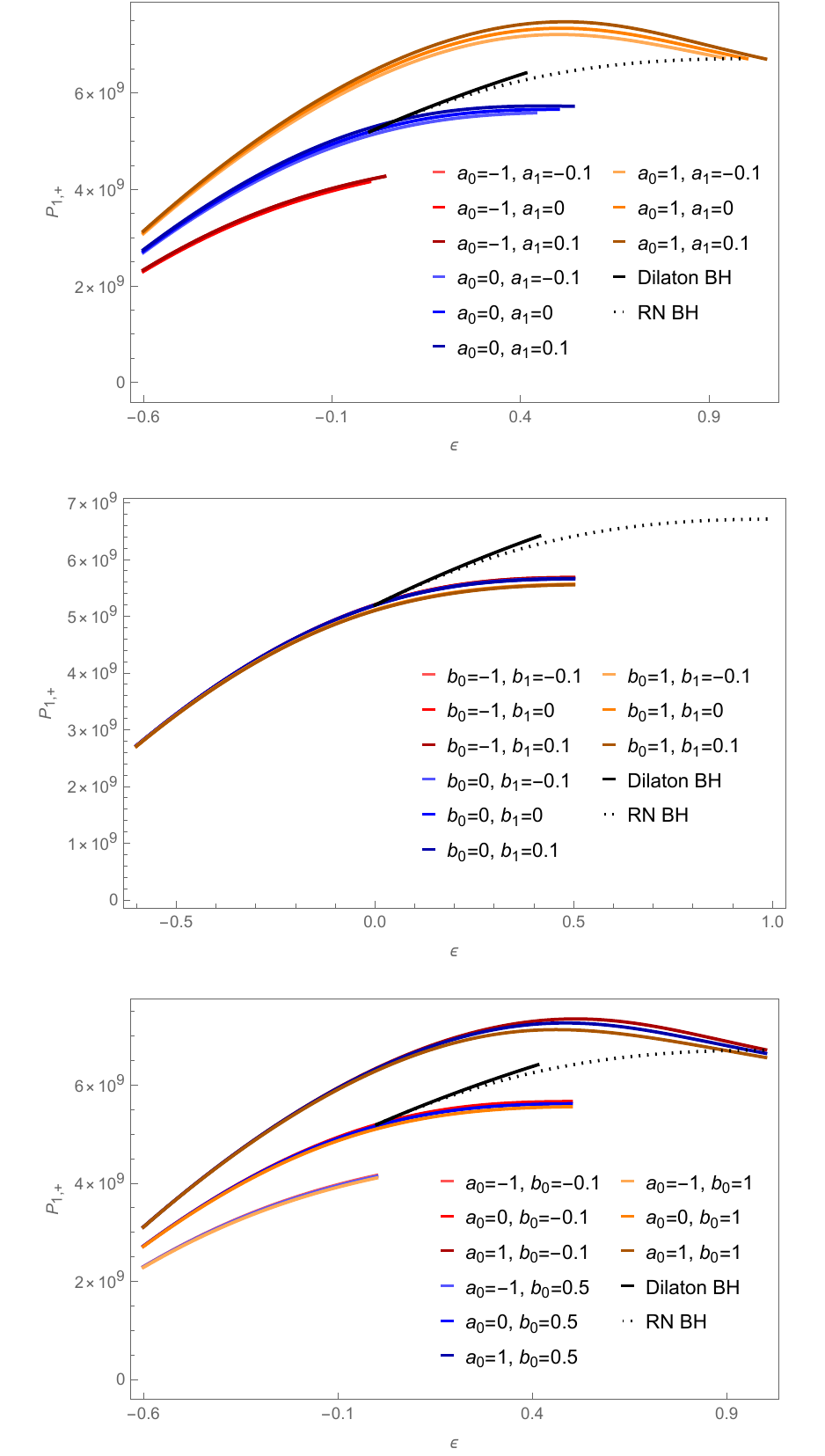}
\centering
\caption{Periodicities of the modulations in the visibility, as a function of $\epsilon$, arising from the interference between the images of order $n = 1$ on opposite sides of the black hole. Different families of curves correspond to different choices of the parameters $a_0$, $a_1$, $b_0$ and $b_1$, as indicated in the legends. The black solid line represents the behavior of the periodicity for the dilaton black hole, while the black dashed line corresponds to that of the Reissner-Nordström black hole. We set the source's azimuthal position, $\varphi_S$, to $45^{\circ}$, $\eta_O = 1$ and $r_S$ to $20M$. In the legend, ``RN BH'' stands for ``Reissner-Nordström black hole''.}
\end{figure}
\section{Perturbative results}

Here, we present analytical results by expanding the various quantities of interest up to first order in the parameters $\epsilon, a_0, b_0, a_1$ and $b_1$, thus elucidating the dominant parameter dependencies and pointing to the general direction of metric-induced effects on observables. The inclusion of higher-order terms is straightforward. By perturbatively solving Eq.\ \eqref{photon sphere equation} and subsequently calculating the minimum impact parameter using Eq.\ \eqref{impact parameter}, one obtains
\begin{align}
    r_m &\simeq 3M - \frac{4}{9}M \left( a_0 + 5\epsilon \right) \coloneqq r_m^{(0)} + r_m^{(1)} \, , \\
    u_m &\simeq 3\sqrt{3}M - \frac{2}{\sqrt{3}}M \left( a_0 + 2\epsilon \right) \coloneqq u_m^{(0)} + u_m^{(1)} \, .
\end{align}
Note that, to first order, $r_m$ and $u_m$ do not depend on the parameter $a_1$. As also specified in Sec.\ II, we also remind the reader that they never depend on the parameters $b_0$ and $b_1$, as seen in Eqs.\ \eqref{photon sphere equation} and \eqref{impact parameter}, respectively.

Using Eq.\ \eqref{SDLcoeffa}, we then calculate the strong deflection limit coefficient $\Bar{a}$, resulting in
\begin{equation}
    \Bar{a} \simeq 1 + \Bar{a}^{(1)} \, ,
\end{equation}
where we defined
\begin{equation}
    \Bar{a}^{(1)} \coloneqq -\frac{2}{27} \left( 4a_0 + 2 a_1 - 9b_0  -6b_1 - 10\epsilon \right) \, .
\end{equation}
As for the coefficient $\Bar{b}$, it involves an integration, as can be deduced from Eq.\ \eqref{coefficient b}. By first expanding the function $g_1(\eta)$ and then performing the integration, the following result is obtained:
\begin{equation}
    \Bar{b} \simeq \Bar{b}^{(0)} + \Bar{b}^{(1)} \, ,
\end{equation}
where we defined
\begin{align}
    \Bar{b}^{(0)} &\coloneqq \log 6 + 2 \sum_{i = S,O} \log \left( \frac{6}{3 + \sqrt{9 - 6 \eta_i}} \right) \, , \\
    \Bar{b}^{(1)} &\coloneqq \Bar{a}^{(1)} \bar{b}^{(0)} + \frac{2}{27} \left[ 6a_0 + 4(a_1 - 3\epsilon) \right] \nonumber \\
    & \hspace{-0.1cm} + \frac{2}{135} \left(c_0 + \sum_{i = S,O} \frac{c_3 \eta_i^3 + c_2 \eta_i^2 + c_1 \eta_i - c_0}{\sqrt{1 - \frac{2\eta_i}{3}}}\right) \, .
\end{align}
In the last of the expressions above, for readability, we introduced the following polynomial expressions:
\begin{align}
    c_0 &\coloneqq 45a_0 + 36a_1 - 15 \left( 9b_0 + 6b_1 + 8\epsilon \right) \, \\
    c_1 &\coloneqq 45a_0 + 42a_1 - 90 b_0 - 90b_1 - 90\epsilon \, , \\
    c_2 &\coloneqq -10a_0 - 18a_1 + 20b_1 + 10\epsilon \, , \\
    c_3 &\coloneqq 4a_1 \, .
\end{align}
The step height is found from Eq.\ \eqref{step height} in Sec.\ \ref{vfcompactsource}. Up to first order, the result is
\begin{align}
    h_{n,+} &= \epsilon_{n,+} = \epsilon_{n,+}^{(0)} + \epsilon_{n,+}^{(1)} \, , \\
    \epsilon_{n,+}^{(0)} &\coloneqq \eta_O \eta_S \, e^{\Bar{b}^{(0)} + \phi_S - (2n + 1)\pi} \, , \\
    \epsilon_{n,+}^{(1)} &\coloneqq \epsilon_{n,+}^{(0)} \left\{ \Bar{b}^{(1)} - \Bar{a}^{(1)} \left[ \Bar{b}^{(0)} + \phi_S \right. \right. \nonumber \\
    & \hspace{3.5cm} \left. \left. - (2n + 1)\pi \right] \right\} \, .
\end{align}
Finally, from Eq.\ \eqref{psameorder} we can obtain the formula for the periodicity of the modulations in the visibility arising from the interference between the images of order $n = 1$ on opposite sides of the black hole. We find
\begin{align}
    P_{n,+} &= P_{n,+}^{(0)} + P_{n,+}^{(1)} \, , \\
    P_{n,+}^{(0)} &\coloneqq \frac{r_O}{u_m^{(0)} \left(2 + \epsilon_{n,+}^{(0)} + \epsilon_{n,-}^{(0)} \right)} \, , \\
    P_{n,+}^{(1)} &\coloneqq -P_{n,+}^{(0)} \left[ u_m^{(1)} + u_m^{(0)} \frac{\epsilon_{n,+}^{(1)} + \epsilon_{n,-}^{(1)}}{2 + \epsilon_{n,+}^{(0)} + \epsilon_{n,-}^{(0)}} \right] \, .
\end{align}
These analytical results reproduce the qualitative behaviors seen in the full numerical results of the previous sections, in the regime of small deviations from Schwarzschild. The analytical approach provides valuable insight into the parametric dependence of the lensing observables, highlighting the role of individual coefficients.

\section{Concluding remarks}

In this paper, we have systematically investigated gravitational lensing phenomena in the strong deflection limit, utilizing the parametrization framework introduced by Rezzolla and Zhidenko \cite{Rezzolla2014}. Through detailed analytical and numerical analyses, we have demonstrated how higher-order images manifest as distinct, measurable signatures in the interferometric visibility function, most notably through a characteristic staircase-like structure. By varying the metric parameters within this framework, we have highlighted clear dependencies between the deviations from Schwarzschild spacetime and the observable interferometric signatures, particularly the step heights and periodicities in the visibility function. These findings significantly contribute to the theoretical groundwork necessary for interpreting data from advanced interferometric observations and potentially enable precise tests of general relativity and the discrimination among alternative gravitational theories.

A natural  future extension of our work is the generalization of the analysis to axially symmetric spacetimes \cite{Konoplya2016}. The complexity introduced by axial symmetry presents additional theoretical and computational challenges; however, overcoming them would greatly enhance our understanding and interpretation of high-resolution astrophysical observations.

Furthermore, we are fully aware that interpreting observational data requires careful distinction between geometric gravitational effects and astrophysical emission processes. As highlighted in Ref.\ \cite{Kocherlakota2022}, neglecting such distinctions can lead to significant degeneracies in interpreting black-hole imaging and interferometric data. Future analyses, therefore, should incorporate detailed models of astrophysical emissions alongside geometric descriptions to provide clearer, more precise tests of gravity theories. Such integrated approaches will ensure a robust extraction of physical information from observations and prevent misinterpretation of data arising from confounding factors. Ultimately, these efforts will leverage upcoming observational breakthroughs to test the limits of general relativity and probe potential deviations from established gravitational paradigms.

\section*{Acknowledgements} 
The authors would like to thank Luciano Rezzolla for his valuable suggestions on a preliminary version of this manuscript.
\label{sec:acknowledgements}

\end{document}